%% file: smc2018template.tex
\documentclass{article}
\usepackage{smc2018}
\usepackage{times}
\usepackage{ifpdf}
\usepackage[english]{babel}
\usepackage{cite}
\usepackage{amsmath,url,times,booktabs, tabularx}
\usepackage{amssymb}

\usepackage{amsmath} 
\usepackage{amssymb} 

\def\papertitle{A SIMPLE FUSION OF DEEP AND SHALLOW LEARNING FOR ACOUSTIC SCENE CLASSIFICATION}
\def\firstauthor{Eduardo Fonseca, Rong Gong, and Xavier Serra}

\newif\ifpdf
\ifx\pdfoutput\relax
\else
   \ifcase\pdfoutput
      \pdffalse
   \else
      \pdftrue
\fi

\ifpdf 
  \usepackage[pdftex,
    pdftitle={\papertitle},
    pdfauthor={\firstauthor},
    bookmarksnumbered, 
    pdfstartview=XYZ 
   ]{hyperref}

  \usepackage[pdftex]{graphicx}
  \graphicspath{{./figures/}}
  \DeclareGraphicsExtensions{.pdf,.jpeg,.png}

  \usepackage[figure,table]{hypcap}

\else 
  \usepackage[dvips,
    bookmarksnumbered, 
    pdfstartview=XYZ 
  ]{hyperref}  

  \usepackage[dvips]{epsfig,graphicx}
  \graphicspath{{./figures/}}
  \DeclareGraphicsExtensions{.eps}

  \usepackage[figure,table]{hypcap}
\fi

\hypersetup{
    colorlinks,%
    citecolor=black,%
    filecolor=black,%
    linkcolor=black,%
    urlcolor=black
}

\title{\papertitle}

\oneauthor
  {\firstauthor} {Music Technology Group, Universitat Pompeu Fabra, Barcelona  \\ %
    {\tt \href{mailto:eduardo.fonseca@upf.edu}{name.surname@upf.edu}}}

\begin{document}
\capstartfalse
\maketitle
\capstarttrue
\begin{abstract}
In the past, Acoustic Scene Classification systems have been based on hand crafting audio features that are input to a classifier. Nowadays, the common trend is to adopt data driven techniques, e.g., deep learning, where audio representations are learned from data. In this paper, we propose a system that consists of a simple fusion of two methods of the aforementioned types: a deep learning approach where log-scaled mel-spectrograms are input to a convolutional neural network, and a feature engineering approach, where a collection of hand-crafted features is input to a gradient boosting machine. We first show that both methods provide complementary information to some extent. Then, we use a simple late fusion strategy to combine both methods. We report classification accuracy of each method individually and the combined system on the TUT Acoustic Scenes 2017 dataset. The proposed fused system outperforms each of the individual methods and attains a classification accuracy of 72.8\% on the evaluation set, improving the baseline system by 11.8\%.\end{abstract}
\section{Introduction}
\label{sec:intro}
\input{S1_Intro}

\section{Method}
\label{sec:method}
\input{S2_Method}

\section{Evaluation}
\label{sec:experi}
\input{S3_Experiments}

\section{Results and Discussion}
\label{sec:results}
\input{S4_Results}

\section{Conclusion}
\label{sec:conclu}

\input{S5_Conclusion}


\bibliography{smc2018bib}

\end{document}

%% file: S1_Intro.tex
Environmental sounds provide contextual information about where we are and the physical events occurring nearby. 
Humans have the ability to identify the environments or the contexts where they are (e.g., park, beach or bus) leveraging only acoustic information.
However, this task is not trivial for systems that attempt to automatize it. One of the goals of \textit{machine listening} is to have systems that perform similarly as humans in tasks like this, which is receiving growing attention from the research community~\cite{virtanen2018computational}.
%


The consecutive editions of the Detection and Classification of Acoustic Scenes and Events (DCASE) Challenges have stimulated the research in this field by benchmarking a variety of approaches for acoustic scene classification and acoustic event detection using common publicly available datasets. 
Acoustic Scene Classification (ASC) can be defined as the task of associating a label to an audio stream thereby identifying the context or environment where the audio stream was recorded, e.g., park or beach~\cite{barchiesi2015acoustic}. 
The acoustic scene consists of all the acoustic material that can be present in a given context, including both background noises and specific acoustic events that may occur either in the foreground or merged as part of the background.
The benefits of machine listening systems performing similarly as humans in recognizing acoustic scenes are manifold. Existing applications range from audio collections management~\cite{landone2007enabling} and intelligent wearable interfaces~\cite{xu2008intelligent} to the development of context-aware applications~\cite{schilit1994context}. 
Some concrete examples include automatic description of multimedia content, or optimization of hearing aids parameter settings based on the recognized scene.

In the past, ASC systems have been based on a \textit{feature engineering} approach, where pre-designed low-level features are extracted from the audio signal and input to a classifier.
The most popular hand-crafted features in audio-related tasks are cepstral features, e.g., MFCCs. Initially taken from the speech recognition field, they are one of the most widespread in ASC too~\cite{aucouturier2007bag,roma2013recurrence}.
Typical examples of classifiers used in ASC include GMM~\cite{aucouturier2007bag} and SVM~\cite{roma2013recurrence}.
The feature engineering approach relies heavily on the capacity of the pre-designed features to capture relevant information from the signal, which in turn may need significant expertise and effort.
In fact, this approach turned out to be neither efficient nor sustainable in many disciplines given the high diversity of problems and particular cases encountered in the real world. 

In recent years, we have witnessed a paradigm shift in ASC similarly to those experienced in areas like computer vision or speech recognition. New techniques have arisen based on \textit{learning representations} from data. These data-driven approaches---especially deep learning---have allowed significant research breakthroughs and have rapidly spread across the audio research community. In this case, the system is able to learn internal representations from a simpler one at the input (typically, a time-frequency representation), and the two stages of the feature engineering approach (feature extraction and classification) are optimized jointly. 
Among the various deep learning approaches available, Convolutional Neural Networks (CNNs) have proved to be effective for several audio related tasks, e.g., speech recognition~\cite{lee2009unsupervised}, automatic music tagging~\cite{dieleman2014end} or environmental sound classification~\cite{salamon2017deep}.
Specifically for ASC, CNNs have also been successfully used e.g.,~\cite{hanconvolutional17,weipingacoustic17}.


%
In this paper, we propose an acoustic scene recognizer that employs the fusion of the two presented trends.
First, a simple 2-layer CNN designed using domain knowledge learns features from mel-spectrograms.
Second, a pool of low-level 
audio features are extracted and input to a Gradient Boosting Machine (GBM).
By combining both approaches with a simple fusion method, we obtain a system that takes advantage of the complementary information that they provide. 
The proposed system is an extension of our previous work~\cite{fonseca2017acoustic}, including improvements in both individual approaches and in the late fusion method, as well as further discussion.
In particular, the main improvements are due to the usage of pre-activation in the CNN, LDA feature reduction in the GBM pipeline, and learning-based late fusion.
The remainder of this paper is organized as follows.
Section~\ref{sec:method} describes the CNN and GBM methods that compose the system.
In Section~\ref{sec:experi} we present the dataset used and the evaluation setup.
Results and discussion for each method individually and the combined system can be found in Section~\ref{sec:results}. 
Section~\ref{sec:conclu} summarizes and concludes this work.


%% file: S2_Method.tex
\subsection{Convolutional Neural Network}
\label{sec:CNN}
\input{S21_CNN}

\subsection{Gradient Boosting Machine}
\label{sec:GBM}
\input{S22_GBM}

\subsection{Late Fusion}
\label{sec:system_latefusion}
In order to combine the predictions from both methods, we tried approaches with and without learning, all of them starting from the individual models' class probabilities computed on the development set using the proposed four-fold cross validation setup.
The simplest approach (i.e., without learning) consists of combining the prediction probabilities by taking their geometric mean, arithmetic mean, or rank averaging. Then, the final predicted label is selected by taking the \textit{argmax} over the resulting values.
The learning-based approach consists of two steps. 
First, using the models' prediction probabilities computed on the \textit{development} set as \textit{training data}, we fit a classifier or \textit{meta learner}. We experimented with logistic regression and SVM with several kernels. The models' hyperparameters were determined by grid search on the training data using four-fold cross validation, trying to restrict the parameter search to ranges providing large regularization.
Then, once the meta learner is fit, we predict labels on the \textit{evaluation} set by taking as input the pre-computed prediction probabilities from CNN and GBM on this set. This approach is sometimes referred to as \textit{stacking}.



%% file: S21_CNN.tex
When CNNs are presented with an audio time-frequency representation, they are able, in theory, to capture spectro-temporal patterns that can be useful for recognition tasks.
Furthermore, the dimensions (width and height) of the convolutional filters can be related to the time and frequency axes, respectively. In this work, we explore how this relation can be exploited when designing the convolutional filters for ASC.


\subsubsection{Audio Pre-processing}
\label{sec:CNN_prepro}
We consider two input representations for the CNN: mel-spectrograms and gammatone-based spectrograms.
Both start with the computation of the power of the short-time Fourier transform (STFT) (using Hamming windows of 40~ms with 50\% overlap) after down-mixing the 2-channel of the binaural files to mono.
In short, the mel-spectrogram aggregates the power values using triangular filters (in the frequency domain) distributed according to the mel scale.
In contrast, the gammatone-based spectrogram aggregates the power values using gammatone filters with center frequencies distributed according to the ERB-rate scale~\cite{wang2006computational}.
For the former we used the Librosa library (v0.5.1), while for the latter we used the Essentia implementation~\cite{Bogdanov_essentia}, which is in turn adapted from~\cite{ellis2009gamma}.
After preliminary experiments we chose mel-spectrograms as input representation, whose computation is detailed next.


A mel filter bank consisting of 128 bands from 0 to 22050 Hz according to Slaney's formula~\cite{slaney1998auditory} is applied to the power of the STFT.
Our mel filter bank presents triangular filters with a peak value of one, as opposed to other filter banks where the filters have equal area.
Finally, the mel energies are logarithmically scaled.
We standardize the log-scaled mel-spectrograms by subtracting the mean and dividing by the standard deviation. We do this on whichever subset of data we use for training. Then, we keep the normalization values and subsequently apply them to standardize the corresponding test set (see Section~\ref{sec:eval_setup}).

Since the recordings of the dataset used are 10s long, the dimensionality of the corresponding spectrograms is considered too high for the proposed architecture. Therefore, they are split into non-overlapping time-frequency patches (T-F patches) or \textit{segments} of 1.5s (i.e., 75 frames). We hence obtain 7 segments per recording, the last one being padded with the last original frame. Thus, the CNN learns from T-F patches of $R^{75\times128}$.


\subsubsection{CNN Architecture}

\begin{figure*}[ht]
    \centering
    \includegraphics[width=\textwidth]{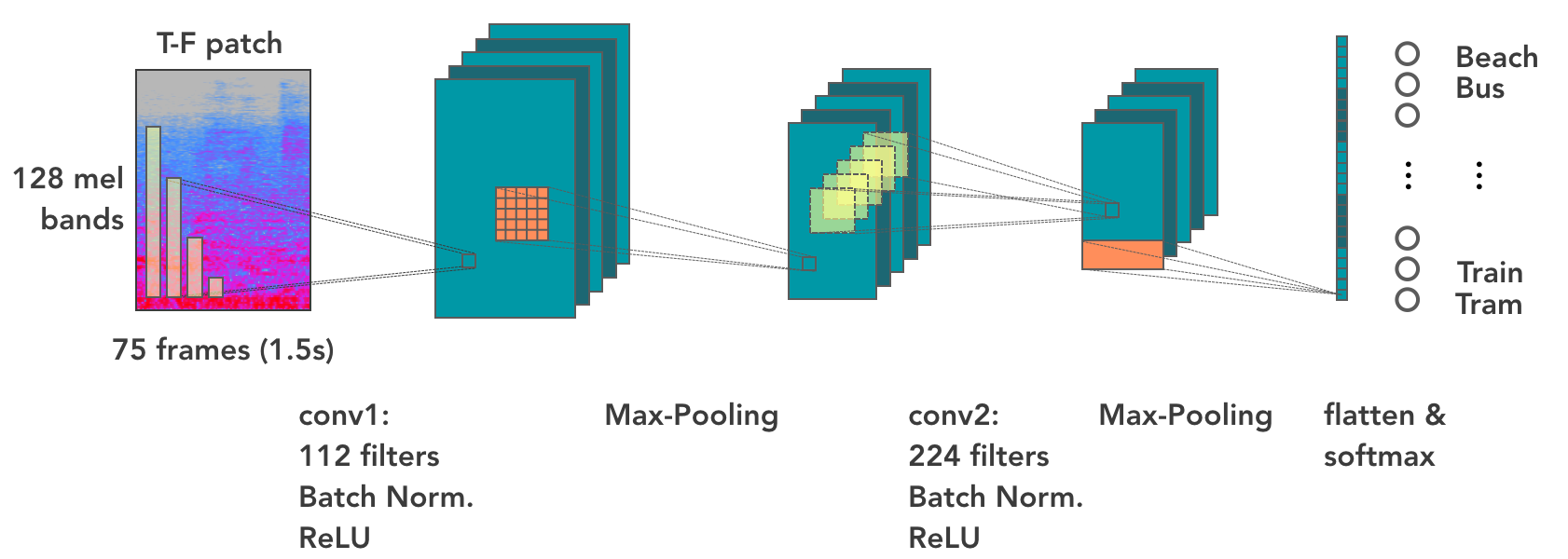}
    \caption{Sketch of the proposed CNN architecture. Four vertical filter shapes co-exist in the first convolutional layer.}
    \label{fig:archi}
    \vspace{-2mm}
\end{figure*}

The proposed CNN architecture is depicted in Table \ref{tab:CNN_arc} and illustrated in Fig. \ref{fig:archi}.

\begin{table}[ht!]
\centering

\begin{tabular}{c}
\hline
Input: 1 x (75,128)           \\
\hline
\textit{Conv1}: 48x (3,8) $|$ 32x (3,32) $|$ 16x (3,64) $|$ 16x (3,90) + \\
BN + ReLU \\
Max-Pooling: (5,5)  	         \\
\hline
\textit{Conv2}: 224x (5,5) + BN + ReLU \\
Max-Pooling: (11,4)  	         \\
\hline
Dense: 15 units + softmax		\\
\hline
\end{tabular}
\caption{Proposed CNN architecture.}
\label{tab:CNN_arc}
\end{table}

The architecture is composed of two convolutional layers (\textit{Conv1} and \textit{Conv2}) alternated with max-pooling operations and it ends with a softmax layer. 
It can be regarded as a relatively simple network comprising standard operations.
Also, the network can be regarded as \textit{wide}, in contrast to the trend of building \textit{deeper} networks with tens of layers (or more in other disciplines like image recognition).

One of the most distinctive aspects of this network is the convolutional filters in the first layer.
We hypothesize that the spectro-temporal patterns that allow to recognize many of the scenes considered are more discriminative along the frequency domain (rather than in the time domain).
We consider this during the filters' design.
That is, our approach attempts to prioritize the modeling of spectral envelope shapes and background noises, rather than onsets/offsets or attack-decay patterns of specific acoustic events. 
While most CNNs in the literature leverage squared filters and only one filter shape in the first convolutional layer~\cite{salamon2017deep,valenti2016dcase,eghbal2016cp}, some recent works suggest to employ rectangular filters and different shapes at the same time~\cite{phan2016robust,pons2017timbre}.
In particular, we explore several configurations of filters with multiple \textit{vertical} shapes in the first layer. We call vertical filters to those whose frequency dimension is much larger than its time dimension.
By using these filters, we intend to aid the learning process towards what we intuitively assume as more important for ASC.

The first convolutional layer is implemented as the concatenation of several convolutional layers such that every layer has filters of one single and distinct shape. Using filters of different dimensions leads to feature maps of different dimensions as well. In order to come back to same-sized feature maps, two options exist: \textit{i)} zero-pad network's input appropriately, and \textit{ii)} use filter-dependent max-pooling operations. Preliminary experiments were run with both options and no major difference in performance was observed. Hence the simpler zero-padding option was adopted.
The filter shapes employed are listed in Table~\ref{tab:CNN_arc} as \textit{number of filters} x \textit{(time, frequency)}. The first convolutional layer presents 112 filters. This number is doubled for the second layer.
The proposed final network presents four different filter shapes in Conv1, as illustrated over the T-F patch of Fig.~\ref{fig:archi}. All the filters in Conv1 have a time dimension of 3. On the contrary, filters in Conv2 are squared 5x5.

We apply batch normalization (BN)~\cite{ioffe2015batch} and Rectified Linear Unit (ReLU)~\cite{glorot2011deep} after every convolutional layer, followed by max-pooling operations. The latter downsample the feature maps while adding some invariance along the time-frequency dimensions. 
In particular, max-pooling is applied over squares of dimension 5 after Conv1.
After Conv2, global time domain pooling is applied in order to select only the most prominent feature~\cite{valenti2016dcase}.
Finally, after flattening the resulting feature maps, the predicted class (for the input T-F patch) is obtained by a dense layer with softmax activation with 15 output units (corresponding to the 15 acoustic scenes).

We also experiment with the concept of \textit{pre-activation}\cite{he2016identity}. This technique was initially devised for image recognition in the context of deep residual networks.
In~\cite{he2016identity} a residual unit is proposed containing two paths: \textit{i)} a clean information path for the information to propagate and \textit{ii)} another path with an additive residual function.
In the latter path, BN and ReLU are applied as pre-activation of the convolutional layers (in addition to the common post-activation consisting of the same couple BN and ReLU after the convolution operation).
Reported advantages in the particular case of deep residual networks, with 100+ layers, include ease of optimization and improved regularization.
Moreover, pre-activation has recently proved successful for ASC in~\cite{hanconvolutional17}, still with a deeper network than the one proposed here. We want to explore this technique in a fairly shallow network.
Based on the results obtained in Section~\ref{sec:CNN_results_PREACT}, we add BN and ReLU non-linearity directly at the network's input of Fig.~\ref{fig:archi} (before the first convolution layer) to form the final proposed CNN.

\subsubsection{Training Strategy and Hyperparameters}
Network weights are initialized with a uniform distribution. 
The loss function is categorical cross-entropy and the optimizer is Adam.
The initial learning rate is 0.002, and it is reduced by a factor of 2 whenever the validation loss does not decrease during 5 epochs. 
We also experimented with \textit{i)} dropping the learning rate by half every fixed number of epochs and \textit{ii)} using Adam with no learning rate scheduling. However, best results were obtained by reducing learning rate when the validation loss plateaus. 
The training is early-stopped if the validation loss is not improved during 15 epochs, up to a maximum of 200. For early-stopping, a 15\% validation set is randomly split from the training data of every class.
The batch size is 64, and training samples are shuffled between epochs.
In both convolutional layers L2 regularization is applied with a parameter of $10^{-5}$.
The system is implemented using Keras (v2.1.3) and Tensorflow (v1.4.1).


%
%

%% file: S22_GBM.tex
Gradient Boosting Machine~\cite{friedman2001greedy} is a technique for constructing predictive models based on an ensemble of many weak learners---typically regression trees. 
The trees are added iteratively, in such a way that the new tree focuses on the misclassifications by the previous ensemble of trees.
Predictions of multiple trees are combined together in order to optimize an objective function, and the parameters of added trees are tuned by gradient descent.
Two GBM frameworks are widely used: XGBoost \cite{xgboost2016} and LightGBM.\footnote{\url{https://github.com/Microsoft/LightGBM}} 
Experiments on five public datasets show that LightGBM outperforms XGBoost on both efficiency and accuracy, with significantly lower memory consumption\cite{GBM_comparison}. In our experiments, we also found out that LightGBM trains faster and achieves a slightly better overall classification accuracy. Hence we use LightGBM in this work.

\subsubsection{Feature Extraction and Pre-processing}
\label{sec:feature_extraction}
We segment each 10s recording into 7 non-overlapped \textit{segments}. 
The first 6 segments last 1.5s, and the last one 1s. 
We then extract features on each segment using the \textit{FreesoundExtractor},\footnote{\url{http://essentia.upf.edu/documentation/freesound_extractor.html}} an out-of-box feature extractor from the Essentia open-source library for audio analysis~\cite{Bogdanov_essentia}.
This extractor computes hundreds of features for sound and music analysis and it is originally used by Freesound\footnote{\url{https://freesound.org/}} in order to provide sound analysis API and searching functionalities.
The most musically-related features (e.g., key, chords, etc.) are discarded.
The selected pool of features is listed in Table~\ref{tab:feature_gbm}, along with their dimensionality. 
The features are calculated at frame-level by using the same frame and hop size mentioned in Section \ref{sec:CNN_prepro}. 
All other parameters of the \textit{FreesoundExtractor} are set to default values. 
We perform four statistical aggregations---mean, variance, and mean and variance of the derivative---to the frame-level feature vectors of each segment. 
Therefore, a $R^{820\times1}$ (i.e., 205$\times$4) feature vector is output for each segment. 
As in Section \ref{sec:CNN_prepro}, we fit a mean and variance standardization scaler over whichever subset of data we use for training, and use it to standardize both train and test data. 

\begin{table}[ht!]
\centering
\label{tab:feature_gbm}
\begin{tabular}{lclc}
\hline
Feature name          & Dim. & Feature name& Dim. \\
\hline
Bark bands energy      	& 32           & Tonal features & 3 \\
ERB bands energy  		& 23           & Pitch features& 3 \\
Mel bands energy        & 45           & Silence rate  & 3\\
MFCC   					& 13           & Spectral features & 32 \\
HPCP      				& 38           & GFCC 	& 13\\

\hline
\end{tabular}
\caption{Selected features extracted by \textit{FreesoundExtractor} and number of dimensions.}
\end{table}

\subsubsection{Linear Discriminant Analysis Feature Reduction}
Linear Discriminant Analysis (LDA) can be used as a dimensionality reduction technique after the feature extraction stage.
The ultimate goal is to mitigate overfitting by projecting a high dimensional dataset onto a lower dimensional space.
This is done by maximizing the variance of the data as well as the separability of classes.
Some of the features of Table~\ref{tab:feature_gbm} are computed in a similar way, e.g., several energy bands are computed with different psychoacoustic scales (e.g., Bark or Mel). 
While they may provide some complementary information, it is likely that they also have a considerable amount of redundancy.
This, together with the high dimensionality of the feature vector, may cause overfitting and a slow-down of model training. 
In order to mitigate this, while keeping the rich information of the extracted features, we perform LDA-based feature reduction.
It is applied on any subset of data used for training, and then the corresponding test set is transformed accordingly (see Section~\ref{sec:eval_setup}).


\subsubsection{Hyperparameter Tuning}
Since ASC is a multi-class classification problem, we use logarithmic loss as the objective function.
We do grid search over 5 hyperparameters. Four of them relate to the GBM (learning rate, \textit{max bins}, \textit{number of leaves}, and \textit{min data in leaf}) while the reduced feature dimension relates to the LDA.
The number of leaves is the main parameter to control model complexity, whereas \textit{max bins} and \textit{min data in leaf} are two important parameters to deal with overfitting. 
All other hyperparameters are set to default values. 
We do the grid search in two cases---with and without LDA---and the hyperparameter values considered are listed in Table~\ref{tab:gridsearch_values}.
The grid search is performed using cross-validation on the development set.
The hyperparameters setting leading to the best cross-validation accuracy is kept for the final GBM model, which is used to predict acoustic scenes on the evaluation set.

\begin{table}[ht!]
\centering
\label{tab:gridsearch_values}
\begin{tabular}{ll}
\hline
Hyperparameter          & Values  \\
\hline
Learning rate      			& [0.01, 0.05, 0.1]            \\
Max bins  					& [128, 256, 512]            \\
Number of leaves        	& [64, 128, 256]           \\
Min data in leaf   			& [500, 1000, 2000]           \\
Reduced feature dimension   & [64, 128, 256, 512]           \\

\hline
\end{tabular}
\caption{Hyperparameter grid search for GBM and LDA.}
\end{table}
\vspace{-2mm}

%% file: S3_Experiments.tex
\subsection{Dataset and Baseline}
\label{sec:dataset}

Systems are evaluated with \textit{TUT Acoustic Scenes 2017}, a dataset that contains recordings made in 15 acoustic scenes. The dataset is split into a development and an evaluation set, of 4680 and 1620 audio recordings respectively.\footnote{A list of the scenes together with more details about the dataset can be found in \url{http://www.cs.tut.fi/sgn/arg/dcase2017/challenge/task-acoustic-scene-classification}.}
The development set contains 312 recordings per class. All recordings last 10s and have a sampling rate of 44.1~kHz. 
A four-fold cross-validation setup is provided for the development set.
The dataset presents a mismatch between development and evaluation set due to differences in the recording conditions. 
The average accuracy drop between both sets across all submitted systems to the ASC task of DCASE2017 is 20.1\%.\footnote{\url{http://www.cs.tut.fi/sgn/arg/dcase2017/challenge/task-acoustic-scene-classification-results}\label{footnote_url_ASC_results}}
A multilayer perceptron (MLP) is provided as baseline system. The input representation is 40 log mel-band energies in 5 consecutive frames and the MLP has 2 layers with 50 hidden units each.

\subsection{Evaluation Setup}
\label{sec:eval_setup}

\begin{figure}[ht]
    \centering
    \includegraphics[width=0.49\textwidth]{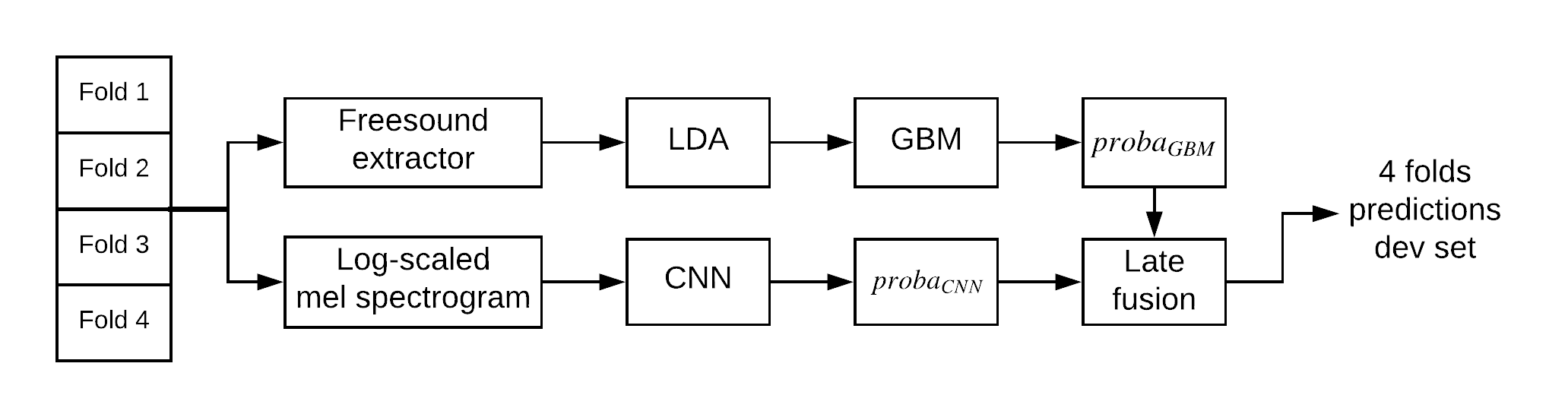}
    \caption{Flowchart illustrating the workflow in development mode.}
    \label{fig:setup_dev}
\end{figure}

The output of CNN and GBM models for every input 1.5s segment is a $R^{15\times1}$ vector with the probabilities of the segment belonging to every class. 
The class prediction for each 10s~recording is computed by averaging per-class scores across segments and finding the class with maximum average score.
The development set is used for training/testing the CNN and GBM approaches according to the provided four-fold cross-validation setup (see Fig.~\ref{fig:setup_dev}).

\begin{figure}[ht]
    \centering
    \includegraphics[width=0.49\textwidth]{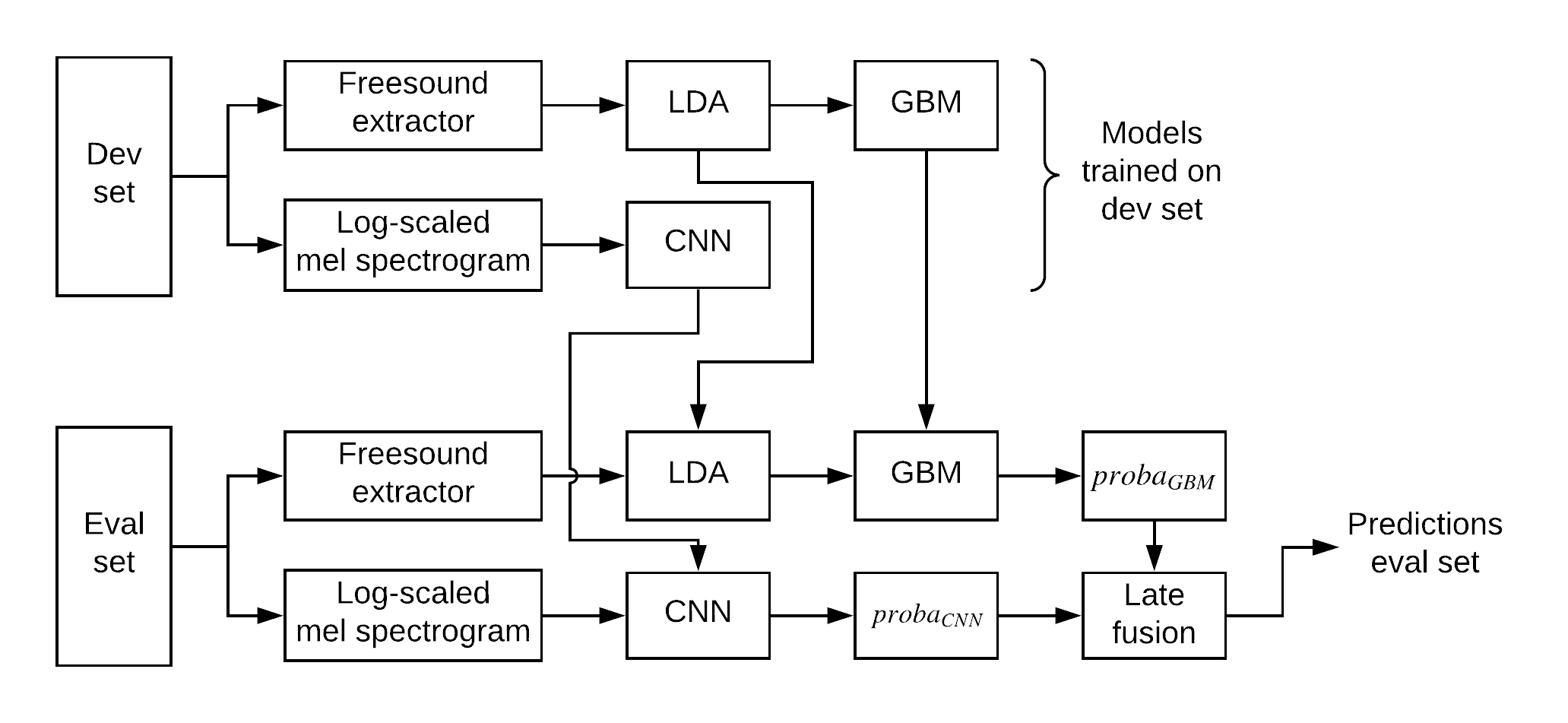}
    \caption{Flowchart illustrating the workflow in evaluation mode. Models are trained on the full development set and predictions are computed on the evaluation set.}
    \label{fig:setup_eval}
\end{figure}

For predicting acoustic scenes on the evaluation set, the models are trained on the full development set and evaluated on the evaluation set (see Fig.~\ref{fig:setup_eval}).
The metric used is classification accuracy, i.e., the number of correctly classified recordings divided by the total amount of recordings.

\vspace{-2mm}


%% file: S4_Results.tex
\subsection{Convolutional Neural Network}
\label{sec:CNN_results}

Two types of experiments were carried out with the CNN: \textit{i)} experimenting with filter configurations in the first layer, and \textit{ii)} exploring the concept of pre-activation.
Since results obtained with GPU are generally non-deterministic, accuracies reported in this Section are the result of averaging ten independent trials of every experiment. Confidence intervals are also shown.

\subsubsection{Filter Configurations}
\label{sec:CNN_results_filters}

\begin{figure}[ht]
\centering
\includegraphics[width=0.52\textwidth]{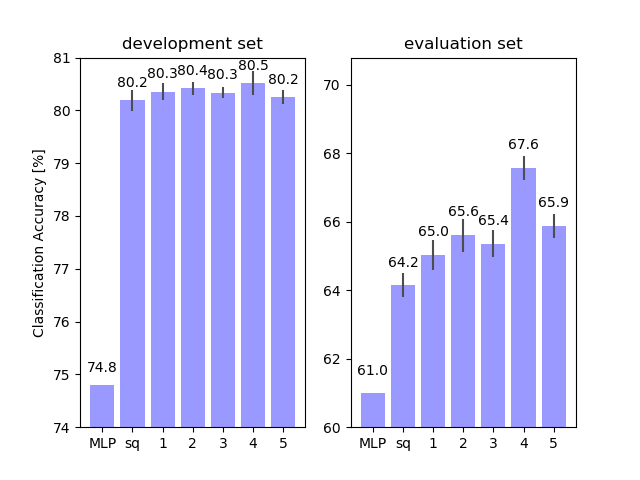}
\vspace*{-8mm}
\caption{ASC performance using the CNN of Fig.~\ref{fig:archi} with the filter configurations in the first layer given by Table~\ref{tab:CNN_filter_conf}. No pre-activation is adopted in these experiments. Note that the y-axis differs for development and evaluation sets.}
\label{fig:cnn_from_sq_to_5}
\end{figure}

We design filter configurations with several filter shapes in the first layer. The number of shapes is specified in Table~\ref{tab:CNN_filter_conf} and Fig.~\ref{fig:cnn_from_sq_to_5} as \textit{CNN\_x}, where $x$ denotes the number of different shapes.\footnote{CNN\_sq refers to the case where filters are squared, which is a specific case of CNN\_1.}
Every shape (denoted by (\textit{time, frequency})) can be repeated a different number of times, as illustrated in Table~\ref{tab:CNN_filter_conf}, but in all cases the total number of filters is 112.

\begin{table}[ht!]
\centering
\label{tab:CNN_filter_conf}
\begin{tabular}{llcc}
\hline
System  		& Filter configuration - \#\textit{filters} x (\textit{time, freq})	\\
\hline
MLP 		      	& -           							 \\
CNN\_sq  			& 112x (5,5)        					\\
CNN\_1    		& 112x (3,40)       						\\
CNN\_2  		& 64x (3,20) $|$ 48x (3,70)    		 		\\
CNN\_3    		& 48x (3,10) $|$ 32x (3,30) $|$ 32x (3,60)   			 \\
CNN\_4			& 48x (3,8) $|$ 32x (3,32) $|$ 16x (3,64) $|$ 16x (3,90)  \\
CNN\_5			& 36x (3,6) $|$ 22x (3,26) $|$ 22x (3,48) $|$ 16x (3,70) \\
				& 16x (3,96) 				  		\\	
\hline
\end{tabular}
\caption{Filter configurations in the first layer for the CNN of Fig.~\ref{fig:archi}.}
\end{table}
\vspace{-1mm}

The motivation for designing filters with different vertical dimensions is to intuitively be able to cover diverse spectral patterns, ranging from narrow-band patterns to others that may spread over frequency.
In order to establish a fair comparison among networks, the number of parameters was kept approximately constant by adjusting the number of filters per shape and the filter dimensions. The number of parameters in all cases lie in the range 656k-660k, with the exception of the squared filters case that has 648k (due to the smaller size of the squared filters). 
In particular, the top performing case of CNN\_4 has 657k parameters. 
The specific filter shapes in Table~\ref{tab:CNN_filter_conf} were chosen through a number of preliminary experiments. While an exhaustive search may be desirable, it may require prohibitively long computation times.

Fig.~\ref{fig:cnn_from_sq_to_5} shows the classification accuracy values for the architecture of Fig.~\ref{fig:archi} and the filter configurations of Table~\ref{tab:CNN_filter_conf}. The accuracy of the MLP baseline is specified as well.
\vspace*{-2mm}

It can be observed that the accuracy on the evaluation set increases overall with the diversity of the filter shapes, until a point where this diversity no longer helps (CNN\_5).
We also carried out some preliminary experiments with horizontal filters but results were slightly worse than those obtained with vertical ones. 




\vspace*{-3mm}
\subsubsection{Pre-activation}
\label{sec:CNN_results_PREACT}

\begin{figure}[ht]
\centering
\includegraphics[width=0.52\textwidth]{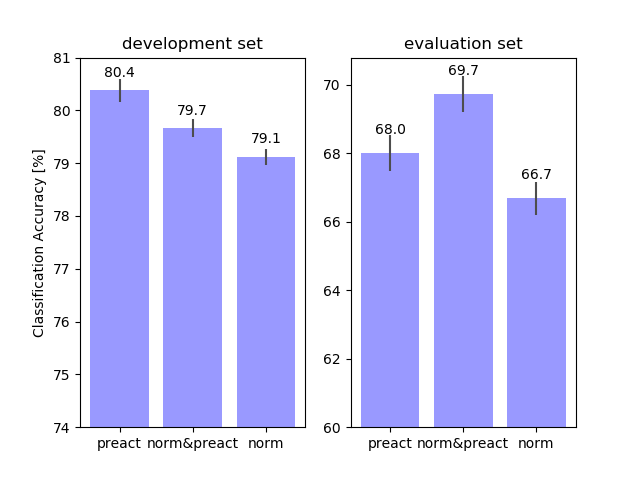}
\vspace*{-8mm}
\caption{ASC performance by adopting pre-activation in the CNN of Fig.~\ref{fig:archi}, i.e., adding BN and ReLU before the first convolutional layer. Note that the y-axis differs for development and evaluation sets.}
\label{fig:cnn_preact_TDnorm}
\end{figure}

Fig.~\ref{fig:cnn_preact_TDnorm} shows the results obtained by adding pre-activation \cite{he2016identity} to the top-performing case of Fig.~\ref{fig:cnn_from_sq_to_5}, i.e., to CNN\_4.
\vspace*{-2mm}

It can be seen that adding pre-activation improves the results slightly on the evaluation set (see \textit{preact} bar). 
However, the gap between development and evaluation accuracies is still substantial. 
Curiously, we found out that this gap is reduced when we complement pre-activation with normalization of the input audio waveform (see \textit{norm\&preact} bar). 
This is somewhat surprising as the T-F patches that input the CNN were already standardized
(see Section~\ref{sec:CNN_prepro}).
Finally, we report the accuracy obtained by applying \textit{only} time domain normalization of audio (without pre-activation), to confirm that it is the combination of both which yields the improvement (see \textit{norm} bar). 
We also experimented with pre-activation not only prior to the first convolutional layer, but also between every max-pooling operation and the next layer, following previous work~\cite{hanconvolutional17}. Resulting accuracies were not higher.

It hence appears that the combination of pre-activation and normalization of the input waveform helps to improve model's generalization, showing slightly lower development accuracy while increasing evaluation accuracy.
Nevertheless, further experiments are needed to better assess and understand the benefits of pre-activation and its dependency on audio signal energy or dynamic range.
For example, one aspect of the audio signal in acoustic scenes or field-recordings is its small dynamic range. 
This happens often as sources can be far away from the microphone, since the goal is to capture the entirety of the acoustic context rather than specific acoustic events. 
Evaluating this approach on different datasets may be revealing.

\subsection{Gradient Boosting Machine}
\label{sec:GBM_results}

The best hyperparameters found for LDA and non-LDA cases 
are listed in Table~\ref{tab:best_values}. 
The dimensionality of the feature vector after LDA-based feature reduction is 64.
This is a 7.8\% of the initial dimensionality (820), which indicates considerable information redundancy in the initial pool of features gathered from the \textit{FreesoundExtractor}. 
After the feature dimension reduction, we observe significant boost in training speed.


\begin{table}[ht!]
\centering
\label{tab:best_values}
\begin{tabular}{lcc}
\hline
Hyperparameter          & non-LDA & LDA  \\
\hline
Learning rate      			& 0.05 & 0.05          \\
Max bins  					& 128  & 128        \\
Number of leaves        	& 128  & 128        \\
Min data in leafs   		& 1000  & 500       \\
Reduced feature dimension   & --   & 64       \\

\hline
\end{tabular}
\caption{Best hyperparameters in both LDA and non-LDA cases by grid searching on the development set.}
\end{table}

Table \ref{tab:gbm_results} shows the accuracy results. 
The performance using LDA feature reduction is greater than the one without LDA and the MLP baseline, resulting in small improvements of 1.7\% and 2.6\% on the evaluation set. 
However, we still witness a significant accuracy drop in both cases.
It is worth to mention that, to tackle the overfitting problem, we have experimented with another two techniques, namely PCA and feature selection using feature importance. However, no significant improvements were observed. 
For the late fusion we use the GBM with LDA.

\begin{table}[ht!]
\centering
\begin{tabular}{lcc}
\hline
Approach          & dev acc (\%) & eval acc (\%) \\
\hline
Baseline			& 74.8		& 61.0		\\
GBM non-LDA  			& 81.4      & 61.9      \\
GBM LDA      			& 81.1      & \textbf{63.6}     	\\

\hline
\end{tabular}
\caption{ASC performance by the GBM model with and without LDA feature reduction.}
\label{tab:gbm_results}
\end{table}

\vspace*{-4mm}
\subsection{Models' Comparison}
\label{sec:comparison}

\begin{figure}[ht]
\centering
\includegraphics[width=0.51\textwidth]{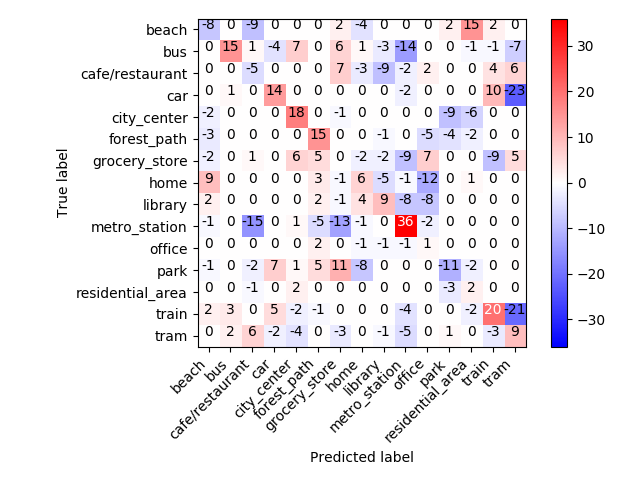}
\vspace*{-6mm}
\caption{Difference between the confusion matrixes produced by \textit{i)} the CNN and \textit{ii)} the GBM models (in this order), evaluated on the evaluation set.}
\label{fig:confusion_diff}
\end{figure}

The CNN method clearly outperforms the GBM method. However, we wanted to assess the potential complementarity of these models, i.e., whether their output predictions are complementary or redundant. We follow the approach of~\cite{salamon2017fusing} consisting of plotting the difference of confusion matrixes yielded by both systems, which is shown in Fig.~\ref{fig:confusion_diff}.

If we have a look at the main diagonal, positive red numbers illustrate scenes where CNN performs better, whereas negative blue numbers represent scenes where the GBM achieves more correct predictions.
The CNN yields better results in most of the acoustic scenes.
However, despite the lower performance of the GBM, it interestingly yields better predictions in the 'park', 'beach' and 'cafe/restaurant' scenes. 
Then, off the diagonal, positive red numbers illustrate that the CNN presents higher confusion between pairs of acoustic scenes. Similarly, negative blue numbers represent that the GBM suffers from higher confusion between pairs of acoustic scenes.
Overall, it can be seen that the models get confused between different pairs of scenes.
In summary, the methods present different behaviour to some extent, and hence their predictions may be complementary.

\subsection{Late Fusion}
\label{sec:fusion_results}

After exploring the approaches described in Section~\ref{sec:system_latefusion}, the logistic regression led to the best results, which are listed in Table~\ref{tab:late_fusion_results}.

\begin{table}[ht!]
\centering
\begin{tabular}{lcc}
\hline
System   							& dev acc (\%)		& eval acc (\%)\\                
\hline
MLP baseline 		   				& 74.8 			& 61.0 \\
Proposed CNN + GBM					& 83.3			& \textbf{72.8}\\
\hline
\end{tabular}
\caption{ASC performance by the combined system.}
\label{tab:late_fusion_results}
\end{table}

\begin{figure}[ht]
    \centering
    \includegraphics[width=0.51\textwidth]{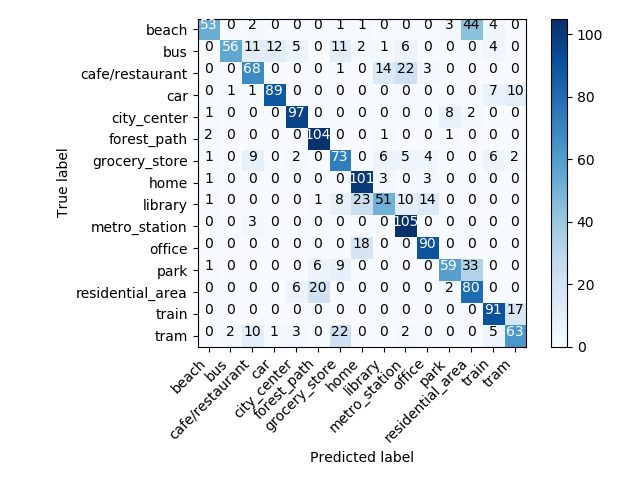}
    \vspace*{-6mm}
    \caption{Confusion matrix for the proposed combined system evaluated on the evaluation set.}
    \label{fig:confusion_fusion}
\end{figure}

The proposed combined system shows an improvement of 3.1\% over the average score provided by the best CNN architecture, and an improvement of 11.8\% over the MLP baseline. It also shows an improvement of 5.5\% with respect to our previous work \cite{fonseca2017acoustic}.
We consider as state of the art the top performing submissions to the ASC task of DCASE2017 Challenge.\textsuperscript{\ref{footnote_url_ASC_results}}
Among them, there are a few systems that outperform the one proposed here.
However, they have the burden of being more complex or computationally intensive, including Generative Adversarial Networks, ensembles of 4 or more systems (with several CNNs), data augmentation, or deeper networks.
Compared to them, we consider that our system is simpler in overall terms. The proposed CNN is more interpretable as domain knowledge was used in its design. The GBM can be trained in a standard desktop computer without need of additional infrastructure, e.g., a GPU.
Figure~\ref{fig:confusion_fusion} shows the confusion matrix for the proposed combined system, where it can be seen which acoustic scenes are misclassified the most. The worst case occurs when the systems predicts 'residential area' while the true label is 'beach' or 'park'.

\vspace*{-1mm}

%% file: S5_Conclusion.tex
We have proposed the fusion of two systems of radically different kind for ASC:
a CNN designed with domain knowledge that learns from log mel spectrograms, and a GBM that leverages audio features from the out-of-box \textit{FreesoundExtractor}.
Evaluated on the \textit{TUT Acoustic Scenes 2017} dataset, the CNN performs substantially better than the GBM, which is not able to generalize well on the evaluation set.
Despite their difference in performance, the models provide somewhat complementary predictions, and their fusion leads to a slight improvement.
The proposed system attains a classification accuracy of 72.8\% on the evaluation set, which means a 11.8\% improvement over the MLP baseline.
Our experiments empirically show that adding pre-activation and waveform normalization help the proposed CNN to reduce overfitting.
Future work includes evaluating the properties of pre-activation on different datasets and networks, and exploring additional measures against overfitting.






\begin{acknowledgments}
This work is partially supported by the European Union's Horizon 2020 research and innovation programme under grant agreement No 688382 ``AudioCommons", and the European Research Council under the European Union's Seventh Framework Program, as part of the CompMusic project (ERC grant agreement 267583), and a Google Faculty Research Award 2017. We are grateful for the GPUs donated by NVidia. 
\end{acknowledgments} 
